\documentclass[twocolumn,superscriptaddress,showpacs,prl,amsmath,amstex,amssymb,citeautoscript,longbibliography]{revtex4-1}
\pdfoutput=1
\usepackage{natbib}
\usepackage[english]{babel}
\usepackage{letltxmacro}
\usepackage{latexsym}
\LetLtxMacro{\ORIGselectlanguage}{\selectlanguage}
\makeatletter
\DeclareRobustCommand{\selectlanguage}[1]{%
  \@ifundefined{alias@\string#1}
    {\ORIGselectlanguage{#1}}
    {\begingroup\edef\x{\endgroup
       \noexpand\ORIGselectlanguage{\@nameuse{alias@#1}}}\x}%
}
\newcommand{\definelanguagealias}[2]{%
  \@namedef{alias@#1}{#2}%
}
\makeatother
\definelanguagealias{en}{english}
\definelanguagealias{English}{english}
\usepackage{graphicx}
\usepackage{amsmath}
\usepackage{amsfonts}
\usepackage{amssymb}
\usepackage{bm}
\usepackage{color}
\usepackage[percent]{overpic}
\usepackage{soul} 
\usepackage{amssymb}
\usepackage{wasysym}
\usepackage{dsfont}
\usepackage{float}

\usepackage{hyperref}
\hypersetup{
    bookmarks=false,         
    unicode=false,          
    pdftoolbar=false,        
    pdfmenubar=true,        
    pdffitwindow=false,     
    pdfstartview={FitH},    
    pdftitle={Bounds on energy absorption in quantum systems with long-range interactions},    
    pdfauthor={W. W. Ho, I. V. Protopopov, D. A. Abanin},     
    pdfsubject={},   
    pdfcreator={},   
    pdfproducer={}, 
    pdfkeywords={many-body localization} {matrix elements} {disordered systems}, 
    pdfnewwindow=true,      
    colorlinks=true,       
    linkcolor=black,          
    citecolor=blue,        
    filecolor=magenta,      
    urlcolor=blue           
}

\newcommand{\be}{\begin{equation}}
\newcommand{\ee}{\end{equation}}
\newcommand{\bea}{\begin{eqnarray}}
\newcommand{\eea}{\end{eqnarray}}

\usepackage{graphicx}
\usepackage[colorinlistoftodos]{todonotes}
\usepackage{verbatim}
\usepackage[normalem]{ulem}

\newcommand{\Tr}{\mathrm{Tr}}

\begin{document}

\title{Bounds on Energy Absorption and Prethermalization in Quantum Systems with Long-Range Interactions
}
\author{Wen Wei Ho}
\affiliation{Department of Theoretical Physics, University of Geneva, 1211 Geneva, Switzerland  }
\affiliation{Department of Physics, Harvard University, Cambridge, Massachusetts 02138, USA}
\author{Ivan Protopopov}
\affiliation{Department of Theoretical Physics, University of Geneva, 1211 Geneva, Switzerland  }
\affiliation{
 L.\ D.\ Landau Institute for Theoretical Physics RAS,
 119334 Moscow, Russia
}
\author{Dmitry A. Abanin}
\affiliation{Department of Theoretical Physics, University of Geneva, 1211 Geneva, Switzerland  }

\date{\today}
\begin{abstract}

Long-range  interacting systems such as nitrogen vacancy centers in diamond and trapped ions serve as experimental setups to probe a range of nonequilibrium many-body phenomena. In particular, via driving, various effective Hamiltonians with physics potentially quite distinct from short-range systems can be realized. In this Letter, we derive general rigorous bounds on the  linear response energy absorption rates of periodically driven  systems  of spins or fermions  with long-range interactions that are sign changing and fall off as $1/r^\alpha$ with $\alpha > d/2$. We show that the disorder averaged energy absorption rate at high temperatures decays exponentially with the driving frequency. This strongly suggests the presence of a prethermal plateau in which dynamics is governed by an effective, static Hamiltonian for long times, and we provide numerical evidence to support such a statement.
Our results are relevant for understanding timescales of  heating and new dynamical regimes described by effective Hamiltonians in such long-range systems.

\end{abstract}

\maketitle

%

\emph{Introduction.} --- Quantum many-body physics far from equilibrium is an exciting frontier of condensed matter physics.
Recent experimental advances in designing well-isolated many-body systems, such as ultracold atoms~\cite{RevModPhys.80.885} and molecules~\cite{1367-2630-11-5-055049}, trapped ions~\cite{Blatt2012} and nitrogen-vacancy (NV) centers in diamond \cite{PhysRevLett.118.093601}, have enabled a controlled study of a range of nonequilibrium phenomena, such as thermalization and many-body localization \cite{RahulReview, AbaninPapic2017, Schreiber842, Chien2015, Monroe16, kucsko,  PhysRevLett.118.093601}.

These experimental platforms generically fall into two classes distinguished by the nature of interactions: short-ranged (e.g.~cold atoms), and long-ranged, power-law decaying (e.g.~NV centers, trapped ions).
Most theoretical work has focused on systems with short-range interactions; 
in contrast, comparatively fewer studies have been conducted on systems with long-range interactions, for which physical phenomena distinct from the former can potentially be realized.
%
%
For example, it was argued that depending on their range, long-ranged interactions can either destroy localization \cite{anderson, Burin06, Levitov:1990zz, Gutman16} or reinstate MBL nonperturbatively \cite{2017arXiv170506290N}.
Furthermore, the existence of a new, critical regime of time crystals was recently uncovered  in a driven dipolar spin system \cite{Choi16DTC, 2017arXiv170304593H}. 
Therefore, studying long-range systems opens up avenues to observe new and interesting physics.

One way to create new dynamical regimes is through periodic driving, which has emerged as a useful tool to engineer interactions
 and create various effective Hamiltonians~\cite{Eckardt2017,Goldman2014, PhysRevLett.111.185301, 2014Natur.515..237J, 2015NatPh..11..162A, 2016arXiv161007611P,PhysRevLett.110.185302}, even allowing for novel nonequilibrium phases of matter such as  time crystals to exist~\cite{Khemani16,Else16,Zhang2017, Choi16DTC}.
However, potential unbounded heating due to the drive can  destroy  such phases, 
\cite{Lazarides14, Ponte14, Alessio14},
and thus it is important to understand the heating timescales in driven many-body systems. 
 Known rigorous results such as exponentially slow heating \cite{PhysRevLett.115.256803} and prethermalization at high driving frequencies \cite{Abanin2017, AbaninPrethermal17,Mori16, Kuwahara201696}, 
however, only apply to systems with sufficiently short-ranged interactions,
and so we would like to understand whether similar general constraints exist in systems with long-range interactions.

In this Letter, we derive general rigorous bounds on the  heating rate  for driven systems of long-range interacting spins (or fermions) in $d$ spatial dimensions at high temperatures. Specifically, we consider  interactions which decay as $1/r^{\alpha}$ with $\alpha > d/2$, and whose coupling strengths are sign-changing and random. We prove  that the {\it disorder-averaged} linear response energy absorption rate is exponentially suppressed at high driving frequencies for both local and global driving.
%
%
This applies to a host of relevant experimental platforms: for example, NV centers interact via long-range dipolar interactions  ($\alpha=d=3$) that are  sign changing in nature; moreover, trapped ion systems can have $\alpha<d$~\cite{Blatt2012}. 
 In order to prove our results, we develop a new method that goes beyond previous works \cite{AbaninPrethermal17,Mori16,Abanin2017, Kuwahara201696} (which relied on the local nature of the interactions). 
We use the fact that it is the random nature of interactions which accords a cancellation of many terms in the response function at high temperatures.
These results strongly suggest the presence of a long-lived prethermal regime described by an  effective, static Hamiltonian, and we support such a statement through numerical studies.

\emph{Setup and results.} --- We consider a many-body system of spins (or fermions) with long-range disordered interactions in $d$ dimensions, placed either on a regular lattice or randomly distributed in space such that there is a short distance cutoff $r_c$, so that the Hamiltonian is
\begin{align}
H = \sum_\mu \frac{J_\mu}{r_\mu^\alpha} O_\mu + \kappa \sum_i \vec{h}_i \cdot \vec{\sigma}_i.
\label{eqn:H}
\end{align}
The sum in $\mu$ is over all links between two sites $(i,j)$ separated by distance $r_{\mu} \equiv r_{ij} \equiv |\vec{r}_i - \vec{r}_j|$, which without loss of generality (WLOG)  is measured relative to 
$r_c$
so that $r_\mu \geq 1$.
$O_{\mu(i,j)} = \sum_{ab} c_{ab} \sigma^a_i \sigma^b_j$ is a generic two-body interaction where $\vec{\sigma}_i$ are Pauli-matrices at site $i$ and $c_{ab}$ fixed real coefficients so that its spectral norm $||O_\mu|| = 1$.
The exponent $\alpha$, characterizing the decay of interactions, is taken to satisfy $d/2<\alpha$, although we are mostly interested in the ``truly'' long-range case for which $d/2 < \alpha \leq d$:
for such   $\alpha$, the mean field on a given site $a$ due to the interactions is not absolutely convergent.
%
We have also allowed for a potentially random on-site field $\sum_i \vec{h}_i \cdot \vec{\sigma}_i$, and assume that $|| \vec{h}_i \cdot \vec{\sigma}_i|| \leq 1$ for all $i$.

We take the interaction strengths $J_{\mu}$ to be independent and identical bounded random variables with 0 mean, so that the following hold for the $n$th moments $J^{(n)} \equiv \langle J^n \rangle$:
\begin{align}
J^{(1)} = 0,  \qquad
J^{(2)} \equiv J^2, \qquad
 | J^{(n)} |  \leq   ( \lambda J)^n, 
\label{eqn:moments}
\end{align} 
for some $\lambda$. 
This is a technical assumption that enables us to derive our bounds: in practice, the interactions of a physical system (such as a dipolar system), while, indeed, sign-changing, are correlated via the relative positions of the spins. However, we believe that our model captures the essential physics of such systems, see \cite{SOM}.
%

We focus on the case of a  harmonic drive of the system at frequency $\omega$ and strength $g$
\begin{align}
H(t) = H + g \cos(\omega t) V,
\label{eqn:drivenH}
\end{align}
where $V = \sum_x V_x $ is a sum of  terms acting on a single site. We take, WLOG, $||V_x|| = 1$ and $\Tr(V_x) = 0$. 
%
%
Assuming the system is initially at thermal equilibrium with inverse temperature $\beta = (k_B T)^{-1}$, the energy absorption rate $dE/dt$ is related to $\sigma(\omega,\beta)$, the dissipative part of the linear response function via $dE/dt = 2g^2 \omega \sigma(\omega,\beta)$. For a quantum system in a finite volume with discrete spectrum, the Lehmann representation of $\sigma(\omega,\beta)$ 
in the high temperature limit $\beta \to 0$ is 
\begin{align}
\sigma(\omega) = \sum_{nm} \frac{\pi \beta \omega }{Z_0} \langle n | V | m \rangle \langle m | V | n \rangle \delta(E_n - E_m - \omega),
\label{eqn:sigma}
\end{align}
where $|n\rangle, |m\rangle$ are energy eigenstates of $H$ and $Z_0 $ is the dimension of the Hilbert space.
 A related quantity was studied  in \cite{PhysRevB.73.035113}. 
Note that Eq.~(\ref{eqn:sigma}) is a  distribution and not a bona fide function -- to state precise results, we have to integrate $\sigma(\omega,\beta)$ over a finite frequency window.
The object of interest  for us is the disordered averaged high-frequency spectral weight of the response function 
$
\sigma([\omega]) \equiv \left\langle \int_\omega^{\infty}  d\omega' \sigma(\omega') \right\rangle,
$
where $\langle \cdot \rangle$ denotes disorder averaging over $J_\mu$ (and, possibly, $h_i$). We derive a bound for $\sigma([\omega])$, establishing the main result of this Letter 
\begin{align}
\sigma([\omega]) \leq  N \pi \beta \omega e^{-\omega/B},
\label{eqn:Bound}
\end{align}
where $N$ is the number of spins in the system and $B > 0$ is some constant that is proportional to the typical two-body interaction strength $J$.
This indicates that the energy absorption of long-range systems at high temperatures is exponentially suppressed at high frequencies. 


\begin{figure}[t]
\center
\includegraphics[width=0.75\columnwidth]{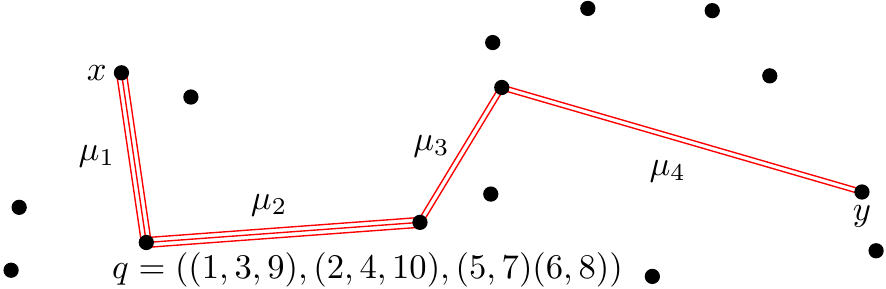}
\caption{A term that contributes in the disorder-averaged $2p$-nested commutator: it is  connected, and all links are at least paired. This particular term corresponds to  $\text{Tr}(V_y [[[ V_x,O_{\mu_1}], O_{\mu_2}], \cdots , O_{\mu_2}]^{(10)})$  with the links appearing in the order  $(\mu_1, \mu_2, \mu_1, \mu_2, \mu_3, \mu_4, \mu_3, \mu_4, \mu_1, \mu_2)$ as given by the partition $q$ in the figure.}
\label{fig:commutators}
\end{figure}

\emph{Sketch of proof: logic and key ideas.}--- The rigorous proof of our result (\ref{eqn:Bound}) is technically heavy, so in this section we simply outline the logic and highlight the key ideas; we refer the reader to \cite{SOM} for  full details. 

The aim is to bound $\sigma([\omega])$ through a careful estimate of the contributions of terms that make up Eq.~(\ref{eqn:sigma}). 
There are three tools we employ. First, under the energy-conserving delta function $\delta(E_n-E_m - \omega)$, the matrix element  $\langle n | V |m \rangle$ can be identically replaced with the $p$-nested commutator $\langle n | [[[V,H],\cdots],H]^{(p)} | m \rangle$, multiplied by a suppression factor $1/\omega^p$.
Since the Hamiltonian $H$ and drive $V$ are sums of at most two-body terms, the $p$-th nested commutator is  a sum of, at most, $(p+1)$th-body terms, each of which is connected (see Fig.~\ref{fig:commutators}).
For example, one such term is $[[[V_x,H_{X_1}], \cdots], H_{X_p}]^{(p)}$, where $H_{X_i}$ is a two-body interaction term making up the Hamiltonian which acts on region $X_i$ --- in order for the commutator not to vanish, regions $x, X_1, \cdots, X_p$ have overlapping support.
%
%
Such a substitution is beneficial, because even though the number of such terms is large, the suppression factor $1/\omega^p$  at high frequencies is small, so this matrix element can be controlled.
The proofs of prethermalization in short-ranged systems \cite{AbaninPrethermal17,Mori16,Abanin2017, Kuwahara201696} essentially relied on this: they explicitly counted the number of connected terms and compared it to the suppression factor to find an exponentially suppressed transition rate; 
%
 however, in our long-range interacting case, we cannot do this because the number of connected terms that appear for any nested commutator is infinite; we thus need a way to both reduce and ``resum'' the individual contributions.

This leads us to our second tool: at high temperatures, all eigenstates contribute, and so we need only consider  the ``matrix element'' of the infinite-temperature ensemble
 $\Tr( V [[[V,H],\cdots],H] )$.
This allows us to, then, apply the third tool, disorder averaging $\langle \cdot \rangle$, independently of the nature of individual eigenstates (which depends on a  particular realization of the Hamiltonian).
The effect of disorder averaging is to kill off many terms in the $p$-nested commutator.
To see this, consider the disorder averaged infinite-temperature matrix element: it is of the form
\begin{align}
\text{Tr}(V[[[V,O_{\mu_1}],O_{\mu_2}],\cdots]^{(p)}) \frac{ \langle J_{\mu_1} J_{\mu_2} \cdots \rangle }{r_{\mu_1}^\alpha r_{\mu_2}^\alpha \cdots}.
\label{eqn:infTempMatrixElement}
\end{align} 
Now because $\langle J_\mu \rangle = 0$, the disorder averaged quantity $\langle J_{\mu_1} J_{\mu_2} \cdots \rangle$ is nonzero only when each link $\mu_i$ appears at least twice, see Fig.~\ref{fig:commutators}. 
In the denominator, distances then come with a power of at least two, i.e.,~$1/r_\mu^{n \alpha}$, $n \geq 2$. 
Anticipating that we will later sum over one of the sites in $\mu = (i,j)$, this higher power guarantees convergence of the sum, i.e.~$\sum_i  {r_{ij}^{-n \alpha}}  < \infty$, as $\alpha > d/2$, so that (the sum of) Eq.~(\ref{eqn:infTempMatrixElement}) is finite. 
Note that  a straightforward bound without the disorder average does not produce a useful result due to the nonabsolutely convergent mean field strength.

Thus, the use of these three tools allows us to reduce the contributions of the infinite number of terms to the linear response function and bound it as Eq.~(\ref{eqn:Bound}).

 \emph{Local drive with no on-site field.}--- 
Let us see how the use of these tools manifestly plays out. 
Consider, as a warmup, proving 
the local version of Eq.~(\ref{eqn:Bound}) for  a local drive, that is, $V = V_0$ acting only on site $0$, and, also, without the on-site field ($\kappa = 0$).
We  rewrite $\sigma([\omega])$ using the first tool of energy conservation as
$$
\sigma([\omega]) =    \left\langle  \int_{\omega}^{\infty} d \omega' \sum_{nm} \frac{\pi \beta \omega'}{Z_0}  \left| \frac{  f^{(p)}_{n,m}}{\omega'^p}  \right|^2 \delta(E_{n} - E_m - \omega') \right \rangle,
$$
where $f^{(p)}_{n,m}  :=   \langle n |  [[[V_0,H],\cdots],H]^{(p)} | m \rangle $.
The integral over $\omega'$ picks out a subset of eigenstates in the double sum, i.e.~eigenstates $|m\rangle$ which differ from $|n \rangle$ in energy by at least $\omega'$. However, we can lift this restriction so that we allow all possible pairs of eigenstates to contribute. This is the infinite temperature ``matrix element'' of the second tool; using the cyclicity of the trace, and various triangle inequalities, we have (see \cite{SOM} for details)
\begin{align}
\sigma([\omega]) 
& \leq   \frac{\pi \beta \omega}{Z_0 \omega^{2p}}
\sum_{\vec{\mu}} \left| \Tr( V_0  [[[V_0, O_{\mu_1}], O_{\mu_2}],\cdots, O_{\mu_{2p}}] )  \right| \nonumber \\
& \times  \frac{\left| \left\langle J_{\mu_1}J_{\mu_2}\cdots J_{\mu_{2p}}\right\rangle \right| }{r_{\mu_1}^\alpha r_{\mu_2}^\alpha \cdots r_{\mu_{2p}}^\alpha }
\label{eqn:Sigma}
\end{align}
 where $\sum_{\vec \mu} = \sum_{\mu_1, \mu_2, \cdots, \mu_{2p} }$, for any $p$. 
This is the form as advertised in Eq.~(\ref{eqn:infTempMatrixElement}). 

We then employ the third tool that, under disorder averaging each $\mu_i$ ($i = 1,2,\cdots,2p$) must be at least paired. 
%
A natural way to account for this in the $2p$-nested commutator is to consider unordered integer partitions $s$ of the integer $2p$ such that each partition is at least two. That is, we denote the set of all integer partitions by
$
S(2p) = \{ s = (n_1, \cdots, n_{l(s)}) ~|~ 
\sum_{k=1}^{l(s)} n_k = 2p ,   n_k \geq 2   \}
$
where  $l(s)$ is the length of the integer partition $s$, so that
$(n_1,\cdots, n_{l(s)})$ corresponds to the number of times the distinct links $(\mu_1, \cdots, \mu_{l(s)})$ appear. 
In addition to specifying the number of times the links appear, we also have to consider different orderings of these $l(s)$ links.
%
To that end, let us for a given integer partition $s$ introduce the set $Q(s)$ of partitions $q$, each $q$ being the list $(1,\cdots, 2p)$ broken into $l(s)$ sublists, such that the length of the $k$th sublist is some  integer part of $s$, see Fig.~\ref{fig:commutators}.
%
We order $q$ by the smallest element appearing in each sublist. 
%
%

With the information $q(s)$, we can specify a connected term in the $2p$-nested commutator of Eq.~(\ref{eqn:Sigma}), namely $f[q(s)] = |\Tr(V_0 [[[V_0, O_{\mu_{i_{1}(q)}}], O_{\mu_{i_{2}(q)}}], \cdots ,O_{\mu_{i_{2p}(q)}}]) |$  where the links $(\mu_1, \mu_2, \cdots, \mu_{l(s)})$ are distributed as follows: $\mu_1$ appears at positions dictated by the first sublist of $q$, $\mu_2$ appears at the positions dictated by the second sublist, and so on.
This allows us to organize and keep track of terms in $\sigma([\omega])$, so that the inequality Eq.~(\ref{eqn:Sigma}) can be expressed identically as
\begin{align}
\sigma([\omega]) \leq &  \frac{\pi \beta \omega}{Z_0 \omega^{2p}} \sum_{s \in S(2p)}  \sum'_{\mu_1,\cdots,\mu_{l(s)}}  \sum_{q \in   Q(s)} f[q(s)]   \times \nonumber \\
&  \frac{|\langle J_{\mu_1}^{n_{1}(q)} \rangle\langle J_{\mu_2}^{n_{2}(q)} \rangle \cdots \langle  J_{\mu_{l(s)} }^{n_{{l(s)} }(q) }\rangle |}{
r_{\mu_1}^{\alpha n_{1}(q)}  r_{\mu_2}^{\alpha n_{2}(q)} \cdots  r_{\mu_{l(s)} }^{\alpha n_{l(s)}(q)   } },
\label{eqn:RHSexact}
\end{align}
where the second sum is over distinct links (denoted by the prime). We see that the effective decay of the interactions has increased under disorder averaging (since $n_k \geq 2$). Note, also, that the numerator can be replaced by a uniform upper bound of $(\lambda J)^{2p}$, c.f.~Eq.~(\ref{eqn:moments}).

Last, all we have to do is carefully estimate the rhs of the above expression  by counting the number of partitions $q(s)$ and integer partitions $s$, as well as summing over tails of the (renormalized) long-range interaction $\sum_{\mu_1, \cdots, \mu_l} r_{\mu_1}^{-\alpha n_1(q)} \cdots  r_{\mu_l}^{-\alpha n_l(q)}$.
 We relegate the detailed analysis to \cite{SOM} and simply quote the result

\begin{align}
\sigma([\omega]) < \pi \beta \omega \left(  2 \sqrt{2 C(2) } \lambda J e^{\nu/2} \right)^{2p}      (2p)! / \omega^{2p},
\label{eqn:LocalBound}
\end{align}
which we see consists of two factors: a suppression term $ \propto (J/\omega)^{2p}$ and a  factorially growing term $(2p)!$ which eventually overcomes the former. $C(2), \nu$ are just numerical factors.
Finding the optimal $p_*$ for which Eq.~(\ref{eqn:LocalBound}) is minimized yields $p_* \propto \omega$, and 
$
\sigma([\omega]) < \pi \beta \omega e^{\omega/B}
$
for a constant $B > 0$  that depends on $J$ and other system size independent numerical factors \cite{SOM}. This is Eq.~(\ref{eqn:Bound}), but without the system size prefactor $N$.

\emph{Global drive.}---   Next, we consider when the drive is global:
 we replace one  $V_0$ term in Eq.~(\ref{eqn:Sigma}) by the global drive $V = \sum_x V_x$ and the other by $V = \sum_y V_y$, and we also have to account for the static on-site field $\sum_i \vec{h}_i \cdot \vec{\sigma}_i$. 
%
%
%
Now, diagonal terms $(x = y)$ simply give rise to contributions already considered in the local case -- this gives a factor $N$ in the bound Eq.~(\ref{eqn:LocalBound}), reflecting the extensivity of the drive, while off-diagonal terms $(x \neq y)$ are additional contributions. 
However, connectivity once again enforces that, for a given $x$, terms arising in the $2p$-nested commutator must have support that overlaps with site $y$ since, otherwise, the commutator vanishes by $\Tr_y(V_y) = 0$; this gives a factor of $p$ more in the bound, which does not affect its scaling. 
Similarly, since the on-site field cannot ``grow'' the support of terms in a $2p$-nested commutator, the growth of the commutator is dominated by the two-body interaction terms, and we have a similar scaling of the bound as before.  Therefore, we obtain our claimed result Eq.~(\ref{eqn:Bound}). We refer the reader to the Supplemental Material \cite{SOM}  for exact details of the derivation.

\emph{Prethermal effective Hamiltonian and numerics.---} Now, let us discuss the implications  of our results. We have shown that 
heating due to direct transitions between eigenstates of $H$ separated by $\omega$ in energy is exponentially suppressed in frequency. While this is a result derived within linear response theory and  at high temperatures, it strongly suggests that there should be a rotating frame of reference [effected by some time-periodic unitary $Q(t)$] in which stroboscopic dynamics is equivalently described by a new Hamiltonian
$
H'(t) \equiv Q(t)^\dagger[ H(t) - i \partial_t ] Q(t) = H_\text{eff} + g_\text{eff} V_\text{eff}(t),
$
such that $H_\text{eff}$ is a {\it static}, effective Hamiltonian, and $V_\text{eff}(t)$ is a remaining  driving piece. Since in this frame $V_\text{eff}(t)$ drives direct transitions between states with energy $\omega$ apart, its effective coupling should be exponentially suppressed, i.e.,~$g_\text{eff} \sim g e^{-\omega/\tilde{J}}$ for some effective interaction strength $\tilde{J}$ which is related to the local energy scale $J$, c.f.~our result Eq.~(\ref{eqn:Bound}). Writing the unitary as $Q(t) = e^\Omega(t)$,  this suggests that  $\Omega(t)$ and so $H_\text{eff}$ are organized in a power series in $(J/\omega)$: $H_\text{eff} = \sum_n^\infty H_n$, where the local norm of $H_n \sim (J/\omega)^n$, see, also, \cite{AbaninPrethermal17,Abanin2017}. $H_\text{eff}$ is then a ``dressed'' version of the undriven Hamiltonian $H$. Hence, dynamics for times $t < t_p \sim g_\text{eff}^{-1}$ before the effects of $V_\text{eff}(t)$ ``kicks in'' should be well captured by the static effective Hamiltonian  $H_\text{eff}$, i.e., a ``prethermal'' regime $(p)$, while for   $t > t_p$, heating due to $V_\text{eff}(t)$ results in a difference  in local observables evolved by the exact Floquet dynamics and the effective Hamiltonian which grows linearly in time:  $\sim t e^{-\omega/\tilde{J}}$. \textcolor{black}{ Note that $H_\text{eff}$ is always a perturbative correction to $H$ as it is obtained at high frequencies; however, this does not preclude its usefulness as this perturbative correction can lead to quite different physics, as in \cite{PhysRevLett.111.185301, 2014Natur.515..237J}.}

\begin{figure}[t]
\center
\includegraphics[width=1\columnwidth]{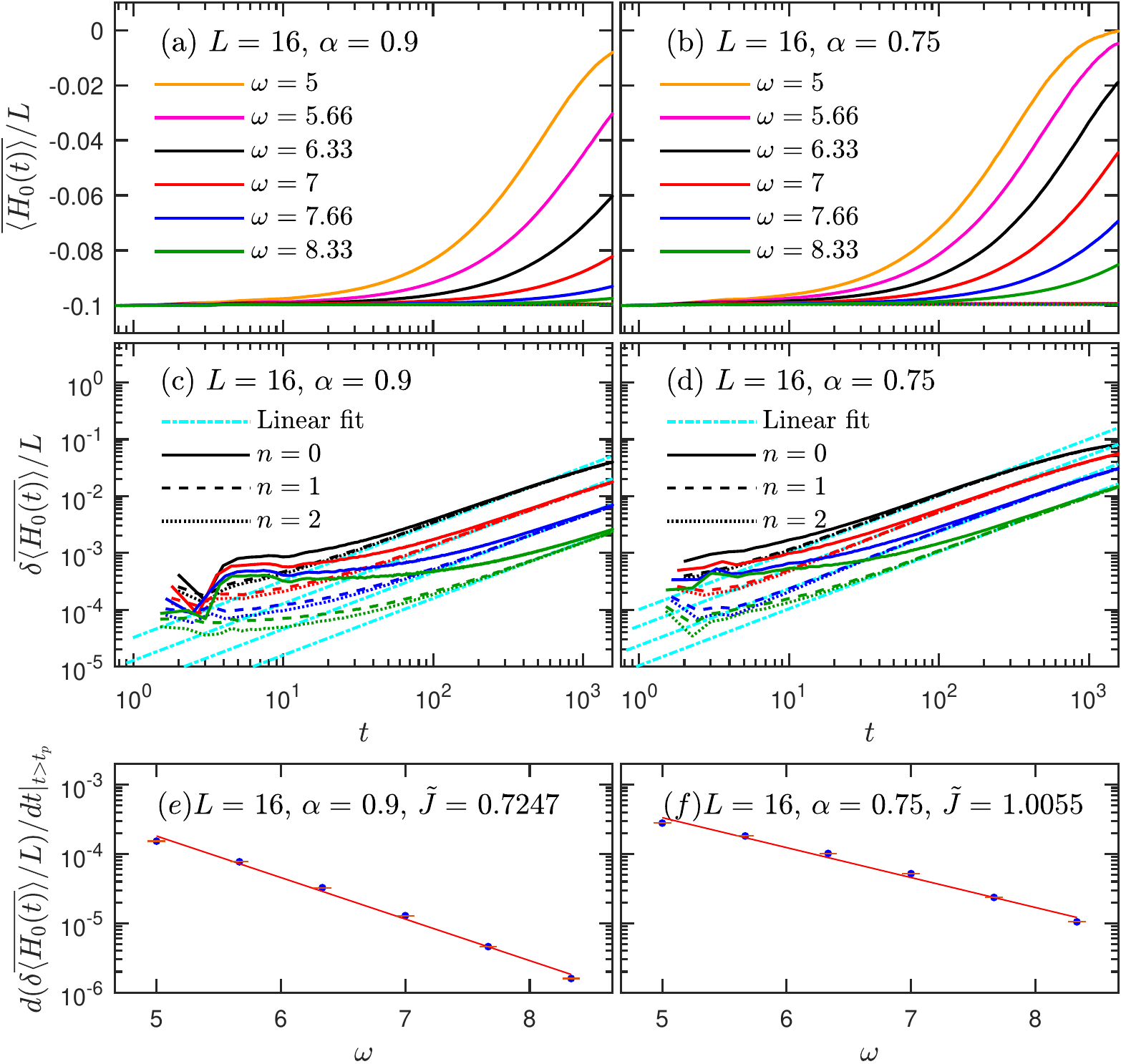}
\caption{Heating in disordered long-range systems with power-law exponents  (a),(c),(e)  $\alpha = 0.9$ and  (b),(d),(f) $\alpha = 0.75$, for various driving frequencies. (a),(b)  Energy density of an initial state evolved by exact Floquet dynamics (solid line) and the $n=2$ effective Hamiltonian (dotted line).  (c),(d) Difference in  energy density between the same initial state evolved by the exact Floquet unitary and $n$-order effective Hamiltonians. Different frequencies are denoted by different colors as in (a),(b).  (e),(f) Rate of (linear) increase of (c),(d)  past the prethermal plateau (cyan linear fit), consistent with scaling $\sim e^{-\omega/\tilde{J}}$, and extraction of effective interaction strength $\tilde{J}$. Results are averaged over 20 disorder realizations (denoted by the bar).}
\label{fig:Numerics}
\end{figure}

 In order to support this appealing picture,  we turn to numerics. We consider a family of 1D long-range spin Hamiltonians  \cite{numericDetails}
\begin{align}
H(t) & = \sum_{ij} \frac{s_{ij}}{r_{ij}^\alpha} ( J_{zz} \sigma_i^z \sigma_j^z + J_{xx} \sigma_i^x \sigma_j^x)  +  \sum_i h_x \sigma_i^x  \nonumber \\
& + g [ 1 - 2 \theta(t - T/2) ] \sum_i (\sigma_i^z + \sigma_i^y),
\label{eqn:numericsH}
\end{align}
where $s_{ij}$ are random in $\{+1,-1\}$ with equal probability  and with $\alpha = 0.9, 0.75$. Because the drive is stepwise, it is natural to utilize the Baker-Campbell-Hauserdoff (BCH) formula and construct a family of effective Hamiltonians $H_\text{eff}^{(n)} \equiv \sum_{k=1}^{n} H_k$ labeled by $n$, defined as the $n$th order truncation of the BCH expansion \cite{SOM}. We note that $H_0$ is nothing but the time-averaged Hamiltonian. 
Initializing a product state in the $z$ basis with energy density $\langle H_0 \rangle/L$ closest to $-0.1$, we  evolve it in time (via Krylov subspace methods) by both exact Floquet dynamics and the $n=0,1,2$ effective Hamiltonians, and measure its energy density $\langle H_0 \rangle/L$.

Figure \ref{fig:Numerics} shows our results. From Figs.~\ref{fig:Numerics}(a) and \ref{fig:Numerics}(b), we see that under, exact Floquet dynamics, the state shows an initial, almost indiscernible very slow heating before an eventual dramatic heating up (to infinite temperature). This initial heating can be attributed to perturbative corrections between effective Hamiltonians. In contrast, it never shows this pronounced heating under the $n=2$ effective Hamiltonians, even at long times. More importantly, characterizing the difference in energy density of the state evolved by exact Floquet dynamics and  higher-order effective Hamiltonians,  Figs.~\ref{fig:Numerics}(c) and \ref{fig:Numerics}(d) show that, at sufficiently high frequencies (larger than $J \sim J_{zz}$ but smaller than the many-body bandwidth), this difference is small and constant for time $t < t_p$, indicating the presence of a ``prethermal'' plateau, while for $t > t_p$  there is a linear increase in the difference, $\delta \overline{\langle H_0(t) \rangle}/L \propto t$.  Extracting the slope of this linear increase in Figs.~\ref{fig:Numerics}(e) and \ref{fig:Numerics}(f) give an excellent agreement  with  $\sim e^{-\omega/\tilde{J}}$, and allows for an extraction of the effective interaction strength $\tilde{J}$. These numerical results support  the presence of effective Hamiltonians in high-frequency driven, disordered, long-range interacting systems.



\emph{Summary and discussion.}--- We have shown that the heating rate of periodically driven, long-range systems is exponentially suppressed at high frequencies, and, furthermore, provided numerial evidence to indicate the presence of a prethermal, effective, static Hamiltonian governing well stroboscopic dynamics for exponentially long times. Thus, this opens up the possibility of realizing new prethermal phases and dynamical regimes, previously discussed only for short-ranged systems.  These results are, in particular, relevant to understanding and constraining dynamics in experimentally accessible setups of long-range interacting degrees of freedom such as dipolar systems realized in ensembles of NV centers in diamond or trapped ions. In the future, it would be interesting to relax the assumption of high temperatures and disorder averaging to prove, nonperturbatively, the existence of a prethermal effective Hamiltonian similar to Refs.~\cite{AbaninPrethermal17,Mori16,Abanin2017, Kuwahara201696}. It would also be appealing to apply our techniques to obtain improved bounds on other dynamical properties in long-range systems, such as entanglement spreading. 
 
\acknowledgements{\emph{Acknowledgments.}---We thank Curt von Keyserlingk, Vedika Khemani, Misha Lukin, Tomotaka Kuwahara and Rahul Nandkishore for useful discussions. We thank the Kavli Institute for Theoretical Physics, where this work was initiated, for hospitality during the program {\it Synthetic Quantum Matter}. W.W.H.~thanks Joonhee Choi for help with the simulations.  This research was supported by the Swiss National Science Foundation, and in part by the Russian Science Foundation under the grant No. 14-42-00044 (IP).
W.W.H.~ is supported by the Gordon and Betty Moore Foundation's EPiQS Initiative through Grant No.~GBMF4306.
}



\bibliography{refs}

\begin{thebibliography}{38}
\bibitem [{num()}]{numericDetails}%
  \BibitemOpen
  \href@noop {} {}\bibinfo {note} {We choose $J_{zz} = 1, J_{xx} = 0.45, h_x =
  0.15, g = 0.1$. Particles are randomly distributed in 1d with PBC with a
  short-distance cut-off of $1$ and average distance of $3$.}\BibitemShut
  {Stop}%
\end{thebibliography}

\clearpage
\section{Supplemental material: Bounds on Energy Absorption and Prethermalization in Quantum Systems with Long-Range Interactions}

\section{Appendix A: Connection of our model to physical models}
In this section, we make a case for why our model, which makes the simplifying assumption that the interaction strengths $J_\mu$ are uncorrelated random variables with $J^{(1)} =0$, $J^{(2)} \equiv J^2$ and $|J^{(n)}| \geq (\lambda J)^n$ for some $\lambda$, captures the essential physics of physical models such as spins interacting via dipolar interactions which are correlated via the relative positions of the spins.

\begin{figure}[t]
\includegraphics[width=1\columnwidth]{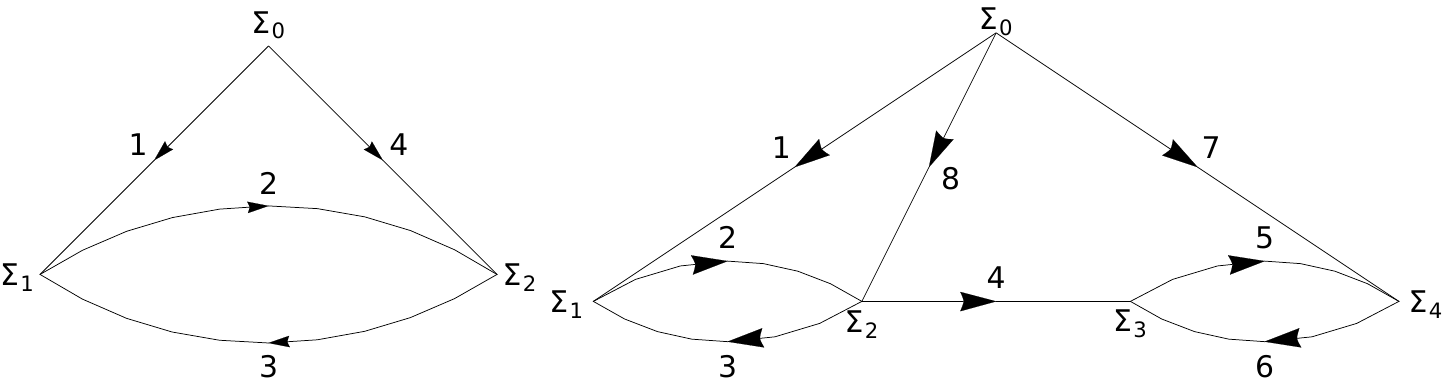} 
\caption{\color{black} Examples of  diagrams which vanish  in the case of uncorrelated disorder, but which survive disorder averaging if the couplings are correlated (for e.g.~spatially). At each vertex is a spin, with the top vertex representing the driven spin $\sigma_0^\alpha$.   Each edge corresponds to a commutation (linking two spins). The arrow represents the edge direction is from the spin that was commuted to the spin that was added.  
The numbers beside the link show the order in which the links appear. 
For example, the left diagram is a graphical representation of the term 
$[[\sigma_0, \sigma_0], \sigma_0][\sigma_1, \sigma_1][\sigma_2, \sigma_2]\sigma_1\sigma_2\rightarrow \sigma_0 \sigma_1\sigma_1\sigma_2\sigma_2 $ appearing in the fourth order commutator
$[[[[V_0, H], H], H], H]$. Here we are only concerned with the number of Pauli matrixes in each vertex and hence we have suppressed both the couplings and the vector indices of the spin operators. 
 } 
\label{simplest}
\end{figure}

On a technical level, we assume that interaction couplings are still sign-changing so that $\langle J \rangle = 0$ but that $J_\mu, J_\nu, J_\rho, \cdots$ are not necessarily uncorrelated, if the links $\mu, \nu, \rho, \cdots$ form a closed loop. This models that the interaction strengths are correlated via the relative positions of the spins, and leads to additional terms in Eq.~(6) of the main text which survive disorder averaging: those that contain links which are not necessarily at least doubled, but which appear as part of a loop. Examples of two such terms are shown graphically in fig.~\ref{simplest}. We would like to check if these terms are divergent and invalidate our bounds, or not.

The crucial feature of the diagram of fig.~\ref{simplest} is that the expression corresponding to a diagram contains at least two Pauli matrixes in each of its vertices (apart from the top vertex where only one Pauli matrix is present). This property is necessary for a diagram to give a nonzero contribution after taking the trace in Eq.~(6) of the main text, since the Pauli matrices are traceless. In our diagrammatic notation, it means that each of the relevant diagrams should have at least two incoming edges in each vertex (apart from the top one). 

Each diagram comes with an integral over the positions of its vertices. 
 For example, suppressing the spin operators and retaining only the spatial dependence, the expressions corresponding to the diagrams of fig.~\ref{simplest} are, for the left figure
\begin{equation}
\int d^d r_1 d^d r_2\frac{1}{r_1^\alpha}\frac{1}{r_2^\alpha}\frac{1}{|r_1-r_2|^{2\alpha}},
\end{equation}
and, for the right figure
\begin{align}
\int d^d r_1 d^d r_2 d^dr_3 d^dr_4 &\frac{1}{r_1^\alpha}\frac{1}{r_2^\alpha}\frac{1}{r_4^\alpha}
\frac{1}{|r_1-r_2|^{2\alpha}} \times \nonumber \\
& \frac{1}{|r_2-r_3|^\alpha}\frac{1}{|r_3-r_4|^{2\alpha}}.
\end{align}

We see that both integrals  are convergent for $\alpha>d/2$. In fact, we have checked \textcolor{black}{for diagrams of the first few} lowest orders (corresponding to orders of commutators) that the above-mentioned requirement for non-vanishing diagrams to have at least two incoming edges per vertex,  renders all the corresponding integrals convergent in the infrared for $\alpha>d/2$ and, particularly so,  in the case of dipolar interactions where $\alpha=d$. Thus, these additional terms which survive disorder averaging are not `dangerous', and hence we do not expect them to modify our bound qualitatively. We conclude that, although the model of uncorrelated disorder studied in our manuscript is a simplification that allows us to prove our rigorous bounds,  we believe it captures correctly the essential physics of physical long-range interacting systems driven at high frequencies.

Note that the same diagrammatic approach can be applied to the ``short-ranged case'', $\alpha>d$ where one  can rigorously prove  that the heating rate is exponentially small. The convergence of the integrals (this is obvious in the ``short-ranged'' case) suggests then that these two systems belong to the same ``universality class'' as long as the high-frequency properties studied in this work are concerned. 
The rigorous proof of this statement requires a careful analysis of the corresponding combinatorics and  
constitutes an interesting direction for future research.

\section{Appendix B: Full details of proof}

In this section we give in full detail the proofs of the bound
\begin{align}
\sigma([\omega]) \leq  N \pi \beta \omega e^{-\omega/B},
\label{eqn:AppendixBound}
\end{align}
presented in the main text. We provide examples to illustrate the counting of the partitions $q(s)$ and integer partitions $s$,
and also more technical estimates for certain constants encountered. 

We will analyze three cases of our long-range driven systems: (i) Local drive with no on-site field, (ii) Global drive with no on-site field, and (iii) Global drive with on-site field. In all three cases, the starting point for the bound on the high frequency part response function is
\begin{align}
\sigma([\omega]) \leq \frac{\pi \beta \omega}{Z_0 \omega^{2p}} \left| \Tr\left( V \left\langle [[[V, H],\cdots],H]^{(2p)} \right\rangle \right) \right|.
\end{align}

To get to this expression, we use the Lehmann representation of the dissipative part of the linear response function $\sigma(\omega)$ and performed the following manipulations:
\begin{widetext}
\begin{align}
\sigma([\omega]) &:=   \left\langle  \int_{\omega}^{\infty} d \omega' \sum_{nm} \frac{\pi \beta \omega'}{Z_0}   \langle n | V | m \rangle  \langle m | V | n \rangle \delta(E_{n} - E_m - \omega') \right \rangle
\nonumber \\
& =  \left\langle  \int_{\omega}^{\infty} d \omega' \sum_{nm} \frac{\pi \beta \omega'}{Z_0}   \frac{    \langle n |  [[[V,H],\cdots],H]^{(p)} | m \rangle \langle m |  [[[V,H],\cdots],H]^{(p)} | n \rangle   }{\omega'^{2p}}   \delta(E_{n} - E_m - \omega') \right \rangle \nonumber \\
& \leq  \frac{\pi \beta \omega}{Z_0 \omega^{2p}}   \left\langle  \int_{\omega}^{\infty} d \omega'  \sum_{nm}   \langle n |  [[[V,H],\cdots],H]^{(p)} | m \rangle \langle m |  [[[V,H],\cdots],H]^{(p)} | n \rangle \delta(E_{n} - E_m - \omega') \right \rangle \nonumber \\
& \leq  \frac{\pi \beta \omega}{Z_0 \omega^{2p}}   \left\langle    \sum_{n} \sum_{m: E_m\in(-\infty, -\omega+E_n)}   \langle n |  [[[V,H],\cdots],H]^{(p)} | m \rangle \langle m |  [[[V,H],\cdots],H]^{(p)} | n \rangle   \right \rangle \nonumber \\
& \leq  \frac{\pi \beta \omega}{Z_0 \omega^{2p}}   \left\langle    \sum_{nm}    \langle n |  [[[V,H],\cdots],H]^{(p)} | m \rangle \langle m |  [[[V,H],\cdots],H]^{(p)} | n \rangle   \right \rangle \nonumber \\
& = \frac{\pi \beta \omega}{Z_0 \omega^{2p}}  \left\langle \Tr\left( [[[V,H],\cdots],H]^{(p)}[[[V,H],\cdots],H]^{(p)} \right)  \right \rangle \nonumber \\
& =  \frac{\pi \beta \omega}{Z_0 \omega^{2p}}  \left|  \left \langle \Tr\left(V [[[V,H],\cdots],H]^{(2p)} \right)  \right\rangle \right| \nonumber \\
& = \frac{\pi \beta \omega}{Z_0 \omega^{2p}}  \left|   \Tr\left(V \left \langle [[[V,H],\cdots],H]^{(2p)}\right\rangle  \right)   \right|,
\end{align}
\end{widetext}
where$|n\rangle, |m\rangle$ are eigenstates of (one disorder realization of) $H$. In the second line, we introduced  $p$ commutators of $H$ with $V$ together with the energy differency $\omega'$, which is an equality under the delta-function; in the third line we uniformly bounded the denominator of all terms by $1/\omega^{(2p)}$ (since every term is positive); in the fourth line we performed the integral and in the fifth line we let the sum extend over all eigenstates. Then, in the second last line, we made use of the fact that $ \left| \Tr([A,H][B,H]) \right| = \left| \Tr([[A,H],H] B) \right|$  repeatedly to transfer all the commutators of one term with $p$-nested commutators to the other term, to end up with a $2p$-nested commutator, while in the last line we made use of the fact the disorder averaging commutes with the trace operation, which is possible only in the high temperature limit. 

\section{Local drive with no on-site field: additional details}
In the case of local driving where $V = V_0$ which is assumed to act only on site $0$ without loss of generality (WLOG), we have
\begin{align}
\sigma([\omega]) 
& \leq   \frac{\pi \beta \omega}{Z_0 \omega^{2p}}
\sum_{\vec{\mu}} \left| \Tr( V_0  [[[V_0, O_{\mu_1}], O_{\mu_2}],\cdots, O_{\mu_{2p}}] )  \right| \nonumber \\
& \times  \frac{\left| \left\langle J_{\mu_1}J_{\mu_2}\cdots J_{\mu_{2p}}\right\rangle \right| }{r_{\mu_1}^\alpha r_{\mu_2}^\alpha \cdots r_{\mu_{2p}}^\alpha }
\label{eqn:AppendixOmega}
\end{align}
 where $\sum_{\vec \mu} = \sum_{\mu_1, \mu_2, \cdots, \mu_{2p} }$ and each sum is over all links in the system. For a finite system of $N$ sites, there are ${N \choose 2}$ links and each sum runs over some enumeration of the links. 

\subsection{Counting using integer partitions $S(2p), S'(2p)$ and partitions $Q(s)$ }

As explained in the main text, because each $\mu_i$ has be to be at least paired (owing to the sign-changing assumption such that $\langle J \rangle = 0$), a natural way to organize the counting is through the number of times distinct $\mu_1, \cdots, \mu_l$ ($l \leq 2p$) appear. This makes us consider the set of all integer partitions of the integer $2p$ with each integer part $\geq 2$, which we denote by $S(2p) = \{ s = (n_1,\cdots n_{l(s)} ) | \sum_{k=1}^{l(s)} n_k = 2p,  n_k \geq 2 \}$, where $l(s)$ is the length of the integer partition. For example, for $2p = 4$, there are two integer partitions $s$:
\begin{align}
2p = 4 = \begin{cases}
 (4) \\
(2,2)
\end{cases}
\end{align}
with lengths $l = 1,2$ respectively. So  Eq.~(\ref{eqn:AppendixOmega}) for the case of a four-nested commutator becomes organized as 
\begin{align}
\sum_{\vec{\mu}} \cdots &= \sum_{\mu_1} \cdots + \sum_{\mu_1} \sum_{\mu_2 \neq \mu_1} \cdots \nonumber \\
& = \sum_{s \in S(2p=4)} \sum_{\mu_1, \cdots, \mu_{l(s)}}' \cdots,
\end{align}
where the prime on the sum means distinct links.

However, as mentioned, we also need to consider the ways that links are distributed in the commutators. This means that for each integer partition $s$, we have to consider {\it its} partitions $q$ of $Q(s)$, which is the set of all partitionings of the list $( 1, 2,3, \cdots ,2p )$ into $l(s)$ sublists, ordered by the smallest element appearing in each part, with the $k$th sublist having $n_k(q)$ elements where $n_k(q)$ is an integer part of $s$. For example, for the integer partition $s = (2,2)$ of $2p = 4$, there are three partitions $q$ of $Q(s)$:
\begin{align}
s = (2,2) \rightarrow Q(s) = \begin{cases} & ((1,2),(3,4)) \\
& ((1,3),(2,4)) \\
& ((1,4), (2,3))
\end{cases}.
\end{align}
As can be seen, each sublist is ordered (and the partition itself is ordered), and the number of elements in each part is this case is $2$. 

What $Q(s)$ gives us information about is   the positionings of how the links are distributed: the first part of a partition tells us the positions in the $2p$-nested commutators where $\mu_1$ appears; the second part of that partition tells us the positions where $\mu_2$ appears, and so on.
For the above example of $2p = 4 $, Eq.~(\ref{eqn:AppendixOmega}) reads 
\begin{align}
& \sum_{\mu_1} |\Tr(V_0 [[[[V_0, O_{\mu_1}],O_{\mu_1}],O_{\mu_1}],O_{\mu_1}]| \frac{| \langle J_{\mu_1}^4 \rangle |}{r_{\mu_1}^{4\alpha} }  + \sum_{\mu_1} \sum_{\mu_2 \neq \mu_1}\nonumber \\
&  \left( |\Tr(V_0 [[[[V_0, O_{\mu_1}],O_{\mu_1}],O_{\mu_2}],O_{\mu_2}]| \frac{| \langle J_{\mu_1}^2 \rangle\langle J_{\mu_2}^2 \rangle |}{r_{\mu_1}^{2\alpha} r_{\mu_2}^{2\alpha} } \right. \nonumber \\
& \left. |\Tr(V_0 [[[[V_0, O_{\mu_1}],O_{\mu_2}],O_{\mu_1}],O_{\mu_2}]| \frac{| \langle J_{\mu_1}^2 \rangle\langle J_{\mu_2}^2 \rangle |}{r_{\mu_1}^{2\alpha} r_{\mu_2}^{2\alpha} } \right. \nonumber \\
& \left. |\Tr(V_0 [[[[V_0, O_{\mu_1}],O_{\mu_2}],O_{\mu_2}],O_{\mu_1}]| \frac{| \langle J_{\mu_1}^2 \rangle\langle J_{\mu_2}^2 \rangle |}{r_{\mu_1}^{2\alpha} r_{\mu_2}^{2\alpha} } \right).
\label{eqn:Appendix2p4Expansion}
\end{align}
More generally, this leads to the exact representation of Eq.~(\ref{eqn:AppendixOmega}) as
\begin{align}
\sigma([\omega]) \leq &  \frac{\pi \beta \omega}{Z_0 \omega^{2p}} \sum_{s \in S(2p)}  \sum'_{\mu_1,\cdots,\mu_{l(s)}}  \sum_{q \in   Q(s)} f(q)   \times \nonumber \\
&  \frac{|\langle J_{\mu_1}^{n_{1}(q)} \rangle\langle J_{\mu_2}^{n_{2}(q)} \rangle \cdots \langle  J_{\mu_{l(s)} }^{n_{{l(s)} }(q) }\rangle |}{
r_{\mu_1}^{\alpha n_{1}(q)}  r_{\mu_2}^{\alpha n_{2}(q)} \cdots  r_{\mu_{l(s)} }^{\alpha n_{l(s)}(q)   } },
\end{align}
 as stated in the main text.

Now, we estimate the r.h.s.~of the above bound by estimating (more precisely, over-estimating the number of integer partitions and partitions present). Let us restrict the sum over integer partitions in $S(2p)$  to be over those that have integer parts being only $2$ or $3$, i.e.~$S'(2p) = \{ s = (n_1,\cdots,n_{l(s)} ) |  \sum_{k=1}^{l(s)} n_k = 2p , 2 \leq n_k \leq 3 \}$, while simultaneously lifting the restriction of distinct links in the sum over $\vec{\mu}$. Then, we get an upper bound 
\begin{align}
\sigma([\omega]) \leq &    \frac{A_p}{Z_0}  \sum_{s \in S'(2p)}  \sum_{\mu_1 \cdots\mu_{l}}  \sum_{q \in  Q(s)}  \frac{  f(q) }{
r_{\mu_1}^{\alpha n_{1}(q)}  \cdots  r_{\mu_{l} }^{\alpha n_{l}(q)   } },
\label{eqn:AppendixUpperbound}
\end{align}
where $A_p = \pi \beta \omega  (\lambda J)^{2p}/ \omega^{2p}$. This is true, because we can cover every integer partition $s$ of $S(2p)$ and its partitions $Q(s)$ (assuming links are distinct) non-uniquely by some integer partition $s$ of $S'(2p)$ together with its partitions $Q(s)$ (assuming some links are distinct). Going back to the example of $2p = 4$, if we  allow $\mu_2$ to run over values of $\mu_1$ in Eq.~(\ref{eqn:Appendix2p4Expansion}), then we can write Eq.~(\ref{eqn:AppendixUpperbound}) as follows. The only integer partition in $S'(2p=4)$ is $(2,2)$, so the bound is 
\begin{widetext}
\begin{align}
& \sum_{\mu_1} \sum_{\mu_2} \left( |\Tr(V_0 [[[[V_0, O_{\mu_1}],O_{\mu_1}],O_{\mu_2}],O_{\mu_2}]| \frac{| \langle J_{\mu_1}^2 \rangle\langle J_{\mu_2}^2 \rangle |}{r_{\mu_1}^{2\alpha} r_{\mu_2}^{2\alpha} } 
 |\Tr(V_0 [[[[V_0, O_{\mu_1}],O_{\mu_2}],O_{\mu_1}],O_{\mu_2}]| \frac{| \langle J_{\mu_1}^2 \rangle\langle J_{\mu_2}^2 \rangle |}{r_{\mu_1}^{2\alpha} r_{\mu_2}^{2\alpha} } \right. \nonumber \\
& \left. |\Tr(V_0 [[[[V_0, O_{\mu_1}],O_{\mu_2}],O_{\mu_2}],O_{\mu_1}]| \frac{| \langle J_{\mu_1}^2 \rangle\langle J_{\mu_2}^2 \rangle |}{r_{\mu_1}^{2\alpha} r_{\mu_2}^{2\alpha} } \right) \nonumber \\
& = 3 \sum_{\mu_1} |\Tr(V_0 [[[[V_0, O_{\mu_1}],O_{\mu_1}],O_{\mu_1}],O_{\mu_1}]| \frac{| \langle J_{\mu_1}^4 \rangle |}{r_{\mu_1}^{4\alpha} }  + \sum_{\mu_1} \sum_{\mu_2 \neq \mu_1}  \left( |\Tr(V_0 [[[[V_0, O_{\mu_1}],O_{\mu_1}],O_{\mu_2}],O_{\mu_2}]| \frac{| \langle J_{\mu_1}^2 \rangle\langle J_{\mu_2}^2 \rangle |}{r_{\mu_1}^{2\alpha} r_{\mu_2}^{2\alpha} } \right. \nonumber \\
& \left. |\Tr(V_0 [[[[V_0, O_{\mu_1}],O_{\mu_2}],O_{\mu_1}],O_{\mu_2}]| \frac{| \langle J_{\mu_1}^2 \rangle\langle J_{\mu_2}^2 \rangle |}{r_{\mu_1}^{2\alpha} r_{\mu_2}^{2\alpha} }  + |\Tr(V_0 [[[[V_0, O_{\mu_1}],O_{\mu_2}],O_{\mu_2}],O_{\mu_1}]| \frac{| \langle J_{\mu_1}^2 \rangle\langle J_{\mu_2}^2 \rangle |}{r_{\mu_1}^{2\alpha} r_{\mu_2}^{2\alpha} } \right)
\end{align}
\end{widetext}
which indeed over-estimates Eq.~(\ref{eqn:Appendix2p4Expansion}). 

Finally, note that we can replace $n_k(q)$ in the exponent of the distances to $2$ since $r_\mu \geq 1$, to get:
\begin{align}
\sigma([\omega]) \leq &    A_p  \sum_{s \in S'(2p)}  \sum_{\mu_1 \cdots\mu_{l}} \frac{1}{r_{\mu_1}^{2\alpha}  \cdots  r_{\mu_{l} }^{2\alpha }  }       \sum_{q \in  Q(s)}  \frac{  f(q) }{Z_0
 }.
\end{align}
We will use this form in the subsequent bounds on $\sigma([\omega])$.

\subsection{ Uniform form on  $\sum_{q \in Q(s)} \frac{f(q)}{Z_0}$}
Let us now prove that
\begin{align}
\sum_{q \in Q(s)} \frac{f(q)}{Z_0} < 2^{2p} \frac{(2p)!}{m_2(s)! m_3(s)!}. 
\end{align}
Here $f(q) = |\Tr(V_0 [[[V_0, O_{\cdot}],\cdots,O_{\cdot} ]]) |$ and the indices on $O$ depend on the partition $q$ in question (and also its connectivity), as discussed before.
$m_{2(3)(s)}$ denote the degeneracies of an integer partition $s \in S'(2p)$'s integer part $2(3)$, i.e.~
\begin{align}
\underbrace{ 2+ \cdots +2}_{m_2(s)} + \underbrace{ 3+\cdots + 3}_{m_3(s)} = 2p.
\end{align}
We use
$$
| \Tr(A) | \leq Z_0 || A ||
$$
where $||\cdot||$ is the spectral norm, which have the following properties:
 $$
|| A B|| \leq ||A || ||B ||, \qquad ||[A,B]|| \leq  2 ||A|| ||B||,
$$
so that 
\begin{align}
\sum_{q \in Q(s)} \frac{f(q)}{Z_0} \leq 2^{2p} \left( \sum_{q \in Q(s)} 1 \right).
\end{align}
The number of partitions in $Q(s)$ is exactly given by
\begin{align}
\frac{(2p)!}{(2!)^{m_2(s)} (3!)^{m_3(s)} } \frac{1}{m_2(s)! m_3(s)!}
\end{align} 
which we upper bound by 
\begin{align}
\frac{(2p)!}{m_2(s)! m_3(s)!},
\end{align}
thereby giving the claimed bound. 

\subsection{Uniform bound on sum over all sites}
Let us now prove that the sum over all sites of the distances (while remembering that links $\mu_1, \mu_2 \cdots, \mu_l$ have to be connected) can be bounded as 
\begin{align}
&\sum_{i_1,\cdots,i_l}  \sum_{a_1 \in \{0,i_1\}} 
 \cdots \sum_{a_{l(s)} \in \{ 0,i_1, \cdots, i_{l(s)-1} \} }  \times \nonumber \\
&  \frac{1}{
r_{0,i_1}^{2\alpha}  r_{a_1,i_2}^{2 \alpha } \cdots  r_{a_{l(s)-1}, i_{l(s)} }^{2\alpha}  }   \nonumber \\
& \leq (m_2(s) + m_3(s))! C(2)^p.
\end{align}
Here $\mu_1 = (0, i_1), \mu_2 = (a_1, i_2), \cdots, \mu_j = (a_{j-1}, i_j)$. 
We perform the summation over $i_1, i_2, \cdots, a_1, a_2,\cdots$ of the distances, but note that $a_{l(s)-1}$ can take at most $l(s) = m_2(s) + m_3(s)$ distinct values, $a_{l(s)-2}$ can take at most $l(s) - 1$ distinct values and so on, and that each sum over $i_k$ for a fixed $a_{k-1}$ is upper bounded by $C(2):= \max_j \sum_i r_{ij}^{-2\alpha}$ which is finite. Then, we can bound the sum over distances by $(m_2(s) + m_3(s))! C(2)^{l(s)} \leq (m_2(s) + m_3(s))! C(2)^p$ since $m_2(s) + m_3(s) \leq p$.

\subsection{Upper bound on number of restricted integer partitions $S'(2p)$}
We estimate the growth of
\begin{align}
\mathcal{N}(2p) = \left( \sum_{s \in S'(2p) } 1 \right).
\end{align}
Since $S'(2p)$ are the integer partitions with integer parts $2$ and/or $3$ only, we have $\mathcal{N}(2p) \leq \frac{p}{3} +1$. For the purposes of the subsequent bound, it will be useful to overestimate this term by an exponential $e^{\nu p}$, the tightest bound for $\nu$ is $\nu = 1/3$, so that
\begin{align}
\mathcal{N}(2p) \leq \frac{p}{3}+1 \leq e^{\nu p}.
\end{align}

\subsection{Optimal $p_*$ and bound on $\sigma([\omega])$}
We find the optimal $p_*$ that minimizes
\begin{align}
\sigma([\omega]) < \pi \beta \omega \left( \frac{2 \sqrt{2 C(2)} \lambda J e^{\nu/2} }{\omega} \right)^{2p} (2p)!.
\end{align}
We use $(2p)! < (2p)^{2p}$, and find that
\begin{align}
p_* = \left\lfloor \frac{\omega}{4 \sqrt{2 C(2)} \lambda J e^{\nu/2 + 1} } \right\rfloor.
\end{align}
At this optimal $p_*$, we then have
\begin{align}
\sigma([\omega]) < \pi \beta \omega e^{-\omega/B},
\end{align}
with $B = 2 \sqrt{2 C(2)} \lambda J e^{\nu/2+1}$.

\section{Global drive with no on-site field: additional details}
In the case of global driving where $V  = \sum_x V_x$ but there is no on-site field, the expression we have to analyze is therefore
\begin{align}
\sigma([\omega]) 
& \leq 
\frac{\pi \beta \omega }{Z_0 \omega^{2p}}
\sum_{x,y}\sum_{\vec{\mu}}   \left| \Tr( V_y  [[V_x, O_{\mu_1}],\cdots], O_{\mu_{2p}}] ) \right| \times \nonumber \\
&     \frac{\left| \left\langle J_{\mu_1}J_{\mu_2}\cdots J_{\mu_{2p}}\right\rangle \right| }{r_{\mu_1}^\alpha r_{\mu_2}^\alpha \cdots r_{\mu_{2p}}^\alpha }.
\end{align}
which is similar to Eq.~(\ref{eqn:AppendixOmega}) except that $x,y$ could act on different sites.

We can account for the off-diagonal terms $(x \neq y)$ via connectivity by introducing delta functions, to get:
\begin{align}
\sigma([\omega]) &\leq A_p   \sum_{x,y}  \sum_{s \in S'(2p)}  \sum_{i_1,\cdots,i_l}  \sum_{a_1 \in \{x,i_1\}} 
\cdots \sum_{a_{l(s)} \in \{ x,i_1, \cdots, i_{l(s)-1} \} } 
 \nonumber \\
&  \sum_{q \in \mathcal Q(s)}  \frac{f_{xy} (q)}{Z_0}   \frac{\delta_{y, i_1} + \delta_{y, i_2} + \cdots \delta_{y, i_{l(s)} } }{
r_{x,i_1}^{\alpha n_{1}(q)}  r_{a_1,i_2}^{\alpha n_{2}(q)} \cdots  r_{a_{l(s)-1}, i_{l(s)} }^{\alpha n_{l(s)}(q)   } },
\label{eqn:GlobalKronecker}
\end{align}
where there are at most $m_2(s) + m_3(s) \leq p$ number of Kronecker-deltas. Performing the sum over $y$ and proceeding as before, we arrive at the following expression to minimize:
\begin{align}
\sigma([\omega]) < N \pi \beta \omega \left( \frac{ 2 \sqrt{2 C(2) } \lambda J e^{\nu/2} }{\omega} \right)^{2p}      (2p)! p,
\end{align}
where $N$ is the total number of sites in the system, reflecting the extensivity of the heating rate. The factor $p$ reflects the maximum number of Kronecker-deltas that arise in enforcing connectivity.

\subsection{Optimal $p_*$ and bound on $\sigma([\omega])$}
We find the optimal $p_*$ that minimizes the previous expression. We use $p \leq e^{p/e}$, so 
\begin{align}
p_* = \left\lfloor \frac{\omega}{4 \sqrt{2 C(2)} \lambda J e^{\nu/2 + 1/e+ 1} } \right\rfloor.
\end{align}
At this optimal $p_*$, we then have
\begin{align}
\sigma([\omega]) < N \pi \beta \omega e^{-\omega/B},
\end{align}
with $B = 2 \sqrt{2 C(2)} \lambda J e^{\nu/2+1/e+1}$.

\section{Global drive with on-site field: additional details}

For the most general case when there is a static on-site field so that $\kappa > 0$, we can write the Hamiltonian  as
$
H = H_{L} + \kappa H_{S} 
$
where $H_L$ is the long-range interaction piece and $H_S = \sum \vec{h}_i \cdot \vec{\sigma}_i$ is the short-range on-site field.
It can be seen somewhat intuitively, from the connectivity analysis done in the preceeding section on the global driving case without on-site field, that the technical analysis and hence the bound in this case will not be changed drastically, because $H_S$ is a sum of on-site terms which can not `grow' the terms in a $2p$-nested commutator. Thus, the growth of the $2p$-nested commutators is completely dominated by the interaction terms, for which we know how to analyze.

The starting point of our analysis is as usual the expression
\begin{align}
\sigma([\omega]) \leq  \pi \beta \omega \frac{1}{\omega^{2p}}  \frac{1}{Z_0}  \left|  \Tr\left(V \left\langle [ [[V,H],H],\cdots,H]^{(2p)} \right\rangle   \right) \right|,
\end{align}
but now because $H = H_L + \kappa H_S$, the $2p$-nested commutator of $V$ with $H$ can be further decomposed into a sum of $2^{2p}$ $2p$-nested commutators of $V$ with  $H_L$ and/or $H_S$, for example $\kappa^2 [[[[V,H_L],H_S],H_S],H_L]$, naturally organized by the number of times $H_L$ (or $H_S$) appears. Denoting by $m_L$ ($m_S$)  the number of times $H_L$ ($H_S$)  appears, we have
$
m_L + m_S = 2p
$
and the number of $2p$-commutators with $H_L$ appearing $m_L$ times is given by the binomial coefficient 
$
{2p \choose m_L }.
$

Let us concentrate on a $2p$-nested commutator with fixed $m_L$ and fixed positions in the $2p$-commutators where $H_S$ acts. Then the analysis proceeds completely analogously as before: ignoring for now the presence of $H_S$, we simply need to consider the integer partitions $s$ of $m_L$ with integer parts being $2$ or $3$,  and consider the partitions $q$ of $s$ with which to distribute the $l(s)$ links, and sum over all such links.  The only difference arises in enforcing connectivity by considering the contributions of $m_S$ $H_S$: for a fixed integer partition $s$ so that
$
2m_2(s) + 3m_3(s) = m_L,
$
 the only potentially non-zero terms due to a commutator with $H_S$ are those terms of $H_S$ (which we recall are a sum of on-site terms) which overlap with the support of the operator produced at the previous level of the $2p$-nested commutator; the number of such sites can be uniformly bounded (regardless of where $H_S$ acts) as $m_L$, so that if there are $m_S$ appearances of $H_S$, there are at most
$
m_L^{m_S} = m_L^{2p - m_L}
$
contributions from all the $H_S$s in this $2p$-nested commutator.

Thus, summing over all possible $m_L$, one gets
\begin{align}
\sigma([\omega]) < & \frac{N \pi \beta \omega  }{\omega^{2p}} \sum_{m_L=0}^{2p} {2p \choose m_L} (\lambda J)^{m_L} \kappa^{2p - m_L} 2^{2p} (m_L)!  \times \nonumber \\
& (2 C(2))^{\frac{m_L}{2}} m_L^{2p - m_L} m_L^2.
\end{align}
The origin of each term in the sum is clear: $(\lambda J)^{m_L} \kappa^{2p-m_L}$ arises from $m_L$ times that $H_L$ appear; $2^{2p}$ arises from the $2p$-nested commutator with all terms having norm $1$; $(m_L)! 2^{\frac{m_L}{2}} C(2)^{\frac{m_L}{2}}$ provides a uniform bound for both the interaction strengths and the number of  partitions $q$ of a given integer partition $s$ of $m_L$; $m_L^{2p - m_L}$ overestimates the contributions from $H_S$ in the $2p$-nested commutator as discussed before, and finally one factor of $m_L$ bounds the number of relevant integer partitions in $S'(m_L)$ and the other factor of $m_L$ bounds the connectivity of a term between $V_x$ and $V_y$.

We can simplify the bound by using $m_L! < m_L^{m_L}$, $m_L^{2p} < (2p)^{2p}$, and $m_L^2 \leq \gamma^{m_L}$ for $\gamma = e^{2/e} $, so that
\begin{align}
& \frac{ \sigma([\omega])}{N \pi \beta \omega} <     \left( \frac{2(2p)}{\omega} \right)^{2p} \sum_{m_L = 0}^{2p} {2p \choose m_L} \left(\lambda J \gamma \sqrt{2 C(2)}\right)^{m_L}  \kappa^{2p - m_L} \nonumber \\
& =   (2p)^{2p} \left( \frac{2 \left(\kappa + \lambda J \gamma \sqrt{2 C(2)} \right)  }{\omega}   \right)^{2p},
\end{align}
recognizing that the sum is a binomial expansion.

\subsection{Optimal $p_*$ and bound on $\sigma([\omega])$}
We find the optimal $p_*$ that minimizes
 the previous expression. We get
\begin{align}
p_* = \left\lfloor \frac{\omega}{4 \left(\kappa +   \lambda J \gamma \sqrt{2 C(2)} \right) e} \right\rfloor.
\end{align}
At this optimal $p_*$, we then have
\begin{align}
\sigma([\omega]) < N \pi \beta \omega e^{-\omega/B},
\end{align}
with $B = 2 \left( \kappa + \lambda J \gamma \sqrt{2 C(2)}  \right) e$. We therefore see that the heating rate is extensive in system size and is likewise exponentially suppressed in the case of global driving with an on-site field.

\section{Baker-Campbell-Hausdorff (BCH) expansion}
We give the expressions for $H_\text{eff}^{(n)} \equiv \sum_{k=1}^n H_k $, which are the $n$-th order truncation of the BCH expansion for our driven long-range spin Hamiltonian used in numerics, for various $n$s.

Let 
\begin{align}
H_0= \sum_{ij} \frac{s_{ij}}{r_{ij}^\alpha} ( J_{zz} \sigma_i^z \sigma_j^z + J_{xx} \sigma_i^x \sigma_j^x)  +  \sum_i h_x \sigma_i^x
\end{align}
 and 
\begin{align}
V = g \sum_i (\sigma_i^z + \sigma_i^y).
\end{align}
 Then 
\begin{align}
& H_1 = i\frac{T}{4}[V,H_0], \\
& H_2 = - \frac{T^2}{24} [V,[V,H_0]].
\end{align}

\end{document}